\documentclass[preprint]{aastex}

\usepackage{psfig}

\newcommand\mzon   {M$_{\odot}$}
\newcommand\pp     {$\pm$}
\newcommand\pers     {s$^{-1}$}

\newcommand\Lunit   {erg s$^{-1}$}
\newcommand\funit   {erg s$^{-1}$ cm$^{-2}$}

\righthead{A super-bursts in 4U 1636-53}
\slugcomment{Accepted for publication in ApJ Letters}

\begin{document}

\title{Recurrent very-long type-I X-ray bursts in the low-mass X-ray
binary 4U 1636--53}

\author{Rudy Wijnands\footnote{Chandra Fellow}}

\affil{Center for Space Research, Massachusetts Institute of
Technology, NE80-6055, 77 Massachusetts Avenue, Cambridge, MA
02139-4307, USA; rudy@space.mit.edu}

\begin{abstract}
Two flares with a duration of several hours are reported for the
low-mass X-ray binary 4U 1636--53. The characteristics of these flares
(i.e., decay time scales, spectral softening, fluences) are very
similar to the very long type-I X-ray bursts recently found in several
other low-mass X-ray binaries, suggesting that the flares in 4U
1636--53 are also very long type-I X-ray bursts. This would make this
source the fifth to exhibit this phenomenon and the first one for
which multiple bursts have been found. Interestingly, all five sources
accrete at approximately 10\% of the Eddington mass accretion
rate. Although a chance coincidence or a selection effect cannot be
ruled-out at present, this correlation is suggestive and might
indicate that only at a narrow range of mass accretion rate such very
long type-I X-ray bursts can occur.
\end{abstract}


\section{Introduction \label{intro}}

Type-I X-ray bursts in low-mass X-ray binary (LMXB) systems are
thought to be due to thermonuclear flashes on the surfaces of neutron
stars, which occur when the unstable burning of helium or hydrogen
ignites the unburned matter on the surface of the neutron stars.
These bursts are characterized by fast rise times (of order seconds),
long decay times (seconds to minutes), spectral softening during the
bursts, and recurrence times of hours to days. Also their X-ray
spectra can usually very well be described by black-body radiation.
The exact physics behind type-I X-ray bursts is very complex, but it
is thought that the properties of the bursts are governed by the
properties of the neutron star (radius, mass, magnetic field strength)
and by the properties of the accreted matter (accretion rate,
composition of the material). For reviews about the X-ray bursts and
the physics involved see, e.g., Lewin et al. (1993) and Bildsten
(1998).

In recent years, exciting new discoveries were reported concerning
type-I X-ray bursts. Soon after the launch of the {\it Rossi X-ray
Timing Explorer} ({\it RXTE}; Bradt, Rothschild, \& Swank 1993) at the
end of 1995, nearly coherent oscillations (called burst oscillations,
with frequencies of 300--600 Hz) were found during the type-I X-ray
bursts of several LMXBs (e.g., Strohmayer et al. 1996). These burst
oscillations are most likely due to the spin frequency of the neutron
star (see Strohmayer 2001 for a review about burst oscillations),
demonstrating that at least several LMXBs harbor rapidly spinning
neutron stars as expected if the neutron star LMXBs are the
progenitors of the millisecond radio pulsars. Cornelisse et al. (2000)
reported an extremely long (hours) type-I X-ray burst (a
``super-burst'') from 4U 1735--44 using the Wide Field Camera (WFC) on
board {\it BeppoSAX}.  Such super-bursts have now also been reported
in several other sources (4U1820--30: Strohmayer 2000; KS 1731--260
and Serpens X-1; Heise, in 't Zand, \& Kuulkers 2000).  The physics
behind such super-bursts is not yet well-known, which is mostly due to
the very recent discovery of such bursts and the limited information
available for them. However, it was suggested that they are not
regular hydrogen or helium flashes, but due to unstable carbon burning
(Strohmayer 2000). Here, I present recurrent super-bursts from the
LMXB 4U 1636--53 as observed with the all sky monitor (ASM; Levine et
al. 1996) on board {\it RXTE}.

\section{The {\itshape RXTE}/ASM light curve of 4U 1636--53}

\begin{figure}[t]
\begin{center}
\begin{tabular}{c}
\psfig{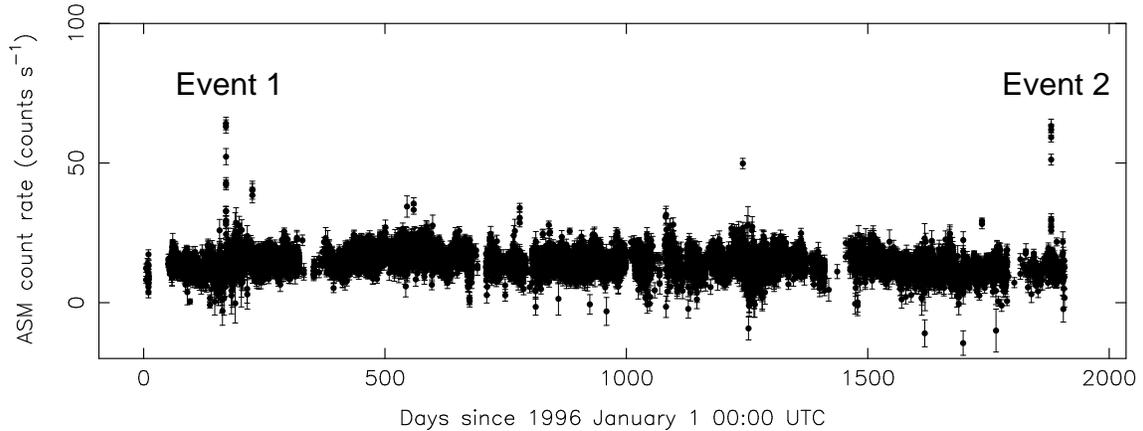}
\end{tabular}
\figcaption{The 1.5--12 keV ASM light curve of 4U 1636--53 showing the
two very long X-ray events.
\label{fig:asm} }
\end{center}
\end{figure}

\begin{figure}[t]
\begin{center}
\begin{tabular}{c}
\psfig{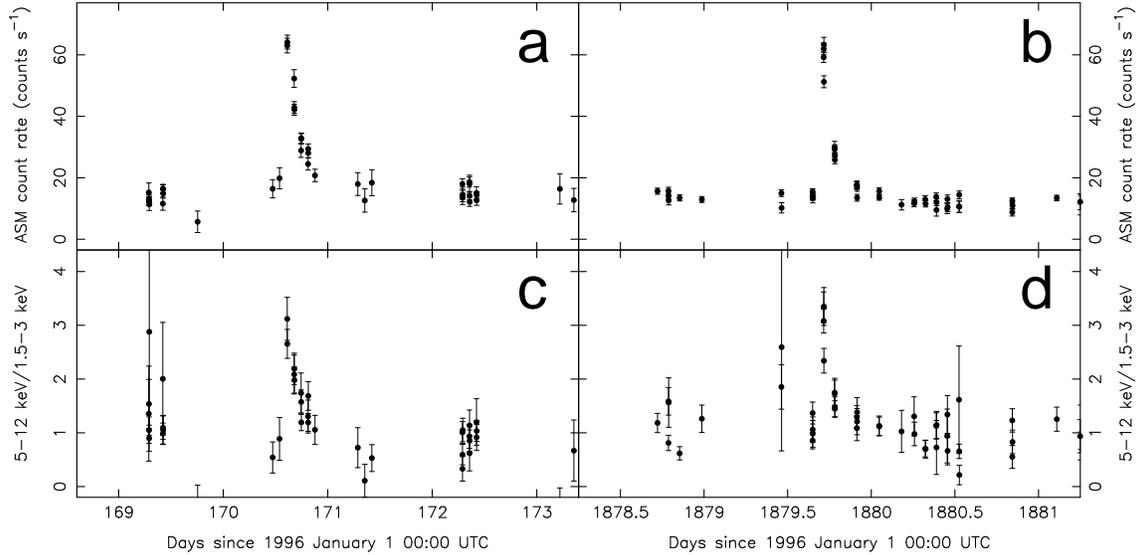}
\end{tabular}
\figcaption{The 1.5--12 keV ASM light curve of 4U 1636--53 for event 1
({\it a}) and 2 ({\it b}), and the 5--12 keV/1.5--3 keV count rate
curve of event 1 ({\it c}) and 2 ({\it d}).
\label{fig:hc} }
\end{center}
\end{figure}

The {\it RXTE}/ASM light curve\footnote{The quick-look {\it RXTE}/ASM
data can be obtained form http://xte.mit.edu/ASM\_lc.html and is
provided by the ASM/{\it RXTE} team} of 4U 1636--53 is shown in
Figure~\ref{fig:asm} (up to 2001 March 22, 13:33 UTC). Clearly visible
is the very recent flare-like event on 2001 Feb 22 17:10 UTC (event
2), just at the end of the light curve. This flare triggered this
study and a close-up of it is shown in Figure~\ref{fig:hc}{\it
b}. Clearly, the rise of this flare is faster than the decay time. The
ratio between the count rate in the range 5--12 keV and the one in the
range 1.5--3 keV is given in Figure~\ref{fig:hc}{\it d}, which shows
that the flux at the peak of the event is harder than the persistent
emission but its spectrum softens considerably during the decay.

As can been seen from Figure~\ref{fig:asm}, another flare-like event
occurred prior to the 2001 Feb 22 one on 1996 June 19 14:42 UTC (event
1), which is shown in detail in Figure~\ref{fig:hc}{\it a}. The event
has a similar peak count rate, profile, and 5--12 keV/1.5--3 keV count
rate ratio (Fig.~\ref{fig:hc}{\it c}) as those of event 2
(Fig.~\ref{fig:hc}{\it d}). These similarities between the two events
strongly suggest that they are due to the same physical mechanism,
demonstrating that this behavior of 4U 1636--53 is recurrent.

The events were fitted with an exponential function in order to
determine a characteristic decay time scale. The results are listed in
Table~\ref{tab:burst}. The first event likely has a longer duration
(about twice as long) as the second event, although the sporadic
sampling of the events with the {\it RXTE}/ASM might have artificially
caused this effect. It is clear, however, that both events lasted more
than one hour and possibly as long as three hours.  The decay time for
event 2 decreased with photon energy (Tab.~\ref{tab:burst}), which is
consistent with the spectral softening of its spectrum during the
decay (Fig.~\ref{fig:hc}{\it b}).  The decay time for event 1
increases with photon energy, although the inverse cannot be excluded
due to the large errors. A smaller decay time at higher energies
(similar to that of event 2) is more likely given the softening of the
spectrum during the decay of event 1.

A detailed spectral analysis of the events is not possible due to the
lack of energy resolution of the {\it RXTE}/ASM data (only three
energy channels). Therefore, the total peak count rates of the events
have been converted into a flux estimate using PIMMS\footnote{A
web-based version of PIMMS is available at
http://heasarc.gsfc.nasa.gov/Tools/w3pimms.html}. As an input
spectrum, a black-body was used with $kT$ of 2 keV (see Cornelisse et
al. 2000 for the temperature of the super-burst in 4U 1735--44, and
assuming the events in 4U 1636--53 are super-bursts as claimed in
section \ref{disc}) and a column density of $4.7\times10^{21}$
cm$^{-2}$ (Schulz 1999; note that due to the relative high energy of
the first energy channel of the {\it RXTE}/ASM [1.5 keV], the flux
results are rather insensitive to the exact column density). The
derived maximum unabsorbed fluxes at the peak of the events are
$2.4\times10^{-8}$ \funit~(1.5--12 keV), which for a distance of 5.9
kpc (Augusteijn et al. 1998) corresponds to a luminosity of
$1.0\times10^{38}$ \Lunit. This is only a lower limit to the peak
luminosity; due to the erratic sampling by the {\it RXTE}/ASM the true
peak might not have been observed.  For the same reason, the total
fluences of the events are rather uncertain, but they are likely to be
in the range 0.5--1 $\times10^{42}$ ergs.

Apart from the two very obvious flares, the {\it RXTE}/ASM light curve
of 4U 1636--53 shows evidence for several smaller flares
(Fig.~\ref{fig:asm}). For these flares, no exponential decays nor a
significant softening of the spectra are observed, although this might
be due to the sampling of the {\it RXTE}/ASM data. It is unclear if
(any of) those smaller flares are due to the same mechanism as the two
large events discussed above. A possible alternative explanation is
that these smaller flares are due to ``normal'' type-I X-ray bursts
which happened to occur at the time that the source was observed with
the {\it RXTE}/ASM.

\section{Discussion \label{disc}}

Two flares were found in the {\it RXTE}/ASM light curve of 4U
1636--53, with characteristic exponential decay profiles which last
for several hours.  The limited energy resolution of the {\it
RXTE}/ASM data does not allow a detailed spectral analysis, but it is
clear that during the decay the flares become considerably softer. The
total peak luminosities and fluences are in the order of $10^{38}$
\Lunit~and $10^{42}$ ergs, respectively. All these characteristics are
very similar to similar flares found with the {\it BeppoSAX}/WFC for
4U 1735--44 (Cornelisse et al. 2000) and the {\it RXTE} proportional
counter array (PCA) for 4U 1820--30 (Strohmayer 2000). These flares
were interpreted as very long type-I X-ray bursts
(``super-burst''). The similarities between those super-bursts and the
flares observed for 4U 1636--53 strongly suggest that the latter
events are also super-bursts. This conclusion is strengthened by the
fact that during the second event, data were also obtained with the
{\it RXTE}/PCA, which again suggests that this flare is a super-burst
(T. Strohmayer 2001, private communication). These high time and
spectral resolution {\it RXTE}/PCA data during this super-burst opens
the exiting possibility of studying a second super-burst in great
detail (after the one already reported for 4U 1820--30; Strohmayer
2000).

4U 1636--53 is the fifth source which exhibits such super-bursts,
demonstrating that this phenomenon is common among the neutron star
LMXBs.  So far, 4U 1636--53 is the only source for which more than one
super-burst has been detected. Two clear super-bursts have been
observed from this source, although some of the smaller flares might
also be related to the super-burst phenomenon. Due to the erratic
sampling of {\it RXTE}/ASM, it is also possible that several other
super-bursts were missed. Therefore, the time span between the two
super-bursts of 4.7 years reflects only an upper limit on the
recurrence time of super-bursts in 4U 1636--53.

Now, with five sources which exhibit these super-bursts identified a
remarkable trend is visible. All five sources are relatively bright,
with a peak intrinsic luminosity of 0.1--0.3 times the Eddington
luminosity ($L_{\rm E}$) for a 1.4 \mzon~accreting neutron star. No
neutron star LMXB with considerably higher ($>$0.5 $L_{\rm E}$) or
lower ($<$0.1 $L_{\rm E}$) peak luminosity has (so far) shown a
super-burst, although this might well be possible due to a chance
coincidence or a selection effect. Due to the higher statistics, the
brightness of the super-burst sources makes it easier to conclusive
identify a possible super-burst with a real super-burst.  However,
this argument is invalid for the brightest neutron star LMXBs, for
which even better statistics are available. Those systems already show
less-normal type-I X-ray bursts, so the mechanism behind the
suppression of the normal bursts might also suppress super-bursts.  A
quick-look of the {\it RXTE}/ASM light curves of the LMXBs currently
monitored by {\it RXTE} did not reveal for those sources an obvious
event like the super-bursts in 4U 1636--53. A more detailed analysis
might reveal less obvious events, but such a study is beyond the scope
of this paper.  A detailed study of the {\it RXTE}/ASM light curves
and/or the {\it BeppoSAX}/WFC of all the burst sources is necessary to
perform a statistical analyses in order to accurately determine how
significant the above noted correlation is.

Although it cannot be excluded that the above correlation is due to
chance coincidence or selection effects, it is also possible that this
correlation is real and that only above a certain (averaged) mass
accretion rate such bursts can occur. If true, then good candidates to
also exhibit super-bursts would be the relatively bright neutron star
LMXBs 4U 1728--34, 4U 1705--44, and 4U 1702--42. The preliminary
examination of their {\it RXTE}/ASM light curve did not reveal a
convincing super-burst. However, this does not mean that they did not
exhibit such burst or can exhibit them in the future, as the
super-bursts reported in 4U 1735--44 and 4U 1820--30 are not present
in the {\it RXTE}/ASM light curves of these sources due to the lack of
coverage by the instrument at the time of the bursts. Furthermore, no
obvious super-bursts are present in the {\it RXTE}/ASM light curves of
KS 1731--260 and Serpens X-1 (the times of the occurrence of the
super-bursts in these sources have not been published yet), although a
possible candidate super-burst occurred in KS 1731--260 at 1996
Sep. 23 12:58 UTC. However, the evidence is inconclusive (i.e., no
evidence for spectral softening is observed) so it is not possible to
claim this event as a super-burst.

The nature and the physics of super-bursts are not well-known, which
is mostly due to the very recent discovery of such bursts and the
limited amount of information available about them. Long type-I X-ray
bursts were already investigated by Fujimoto et al. (1987), but they
predicted that such very long bursts would only occur at an accretion
rate below 1\% of the critical Eddington rate. However, as already
pointed-out for 4U 1735--44 by Cornelisse et al. (2000), all five
sources exhibiting super-bursts accrete well above this level. This
discrepancy might be due to the fact that Fujimoto et al. (1987) only
considered hydrogen and helium flashes and not carbon flashes, which
Strohmayer (2000) suggests as a mechanism for the super-burst in 4U
1820--30. He found that just prior to the super-burst in 4U 1820--30,
a regular helium flash occurred, demonstrating that a helium flash
nature for the super-burst in this source can be excluded, suggesting
that the super-burst might be due to a carbon flash.  Taam \& Picklum
(1978) studied carbon flashes and found that such bursts occur for
accretion rates between $10^{-10}$ and $10^{-9}$ \mzon~yr$^{-1}$,
which is only slightly lower than the accretion rates observed for the
five super-burst sources (see also Strohmayer 2000).  

However, Brown \& Bildsten argued that only for pure helium accreting
sources, enough carbon is expected to remain available after normal
type-I bursts, that a carbon burst could occur. Therefore, carbon
bursts could explain the super-burst in 4U 1820--30, which has a white
dwarf companion, but it less likely that they can explain the
super-bursts in 4U 1636--53 for which hydrogen has been observed in
its optical spectrum (see, e.g., Canizares, McClintock, \& Grindlay
1979; Augusteijn et al. 1998). More theoretical investigations are
needed to determine if all super-bursts are due to carbon flashes or
that a different mechanism is necessary for the super-bursts in
certain sources.

Instruments like the {\it BeppoSAX}/WFC and the {\it RXTE}/ASM are
good to study many sources simultaneously to increase the chances of
observing a super-burst, but they have limited capabilities (limited
spectral and timing resolution and/or limited sensitivity). In order
to get a better understanding of super-bursts, high resolution
spectral instruments and/or high timing resolution instruments are
necessary.  However, due to the rarity of super-bursts such data are
difficult to obtain. It is very fortunate that such data exist now,
not only for the super-burst in 4U 1820--30 (Strohmayer 2000), but
also for one in 4U 1636--53 (T. Strohmayer 2001 private
communication).

\acknowledgments

This work was supported by NASA through Chandra Postdoctoral
Fellowship grant number PF9-10010 awarded by CXC, which is operated by
SAO for NASA under contract NAS8-39073.  This research has made use of
the quick-look results provided by the ASM/{\it RXTE} team.  I thank
Jon Miller, Tod Strohmayer, and Remon Cornelisse for helpful and
insightful discussions.

\begin{deluxetable}{ccccc}
\tablecolumns{5}
\tablewidth{0pt}
\tablecaption{Event parameters \label{tab:burst}}
\tablehead{Time of event$^a$ & Peak count rate$^b$  &
\multicolumn{3}{c}{Characteristic decay time scale (min)} \\\cline{3-5}
            (UTC)      & (ASM counts \pers) & 1.5--12 keV & 1.5--5 keV
& 5--12 keV}
\startdata
1996 Jun 19 14:42 & 64\pp2 & 187$^{+35}_{-29}$ & 158$^{+130}_{-72}$& 187\pp29\\
2001 Feb 22 17:10 & 62\pp2 &  88\pp7           & 132$^{+22}_{-19}$ & 68\pp6\\
\enddata
\tablenotetext{a}{Time of the peak of the events}
\tablenotetext{b}{The total ASM count rate at the peak of the events
for the 1.5--12 keV energy range}
\end{deluxetable}

\end{document}